\documentclass[conference]{IEEEtran}
\IEEEoverridecommandlockouts

\usepackage{cite}
\usepackage{amsmath,amssymb,amsfonts}
\usepackage{algorithmic}
\usepackage{graphicx}
\usepackage{textcomp}
\usepackage{xcolor}
\def\BibTeX{{\rm B\kern-.05em{\sc i\kern-.025em b}\kern-.08em
    T\kern-.1667em\lower.7ex\hbox{E}\kern-.125emX}}
\begin{document}

\title{A Framework for Adaptive Multi-Turn Jailbreak Attacks on Large Language Models
}

\author{
\IEEEauthorblockN{%
  Sidhant Narula\IEEEauthorrefmark{1}\textsuperscript{1},
  Javad Rafiei Asl\IEEEauthorrefmark{1}\textsuperscript{1},
  Mohammad Ghasemigol\IEEEauthorrefmark{1},
  Eduardo Blanco\IEEEauthorrefmark{2},
  Daniel Takabi\IEEEauthorrefmark{1}}
\IEEEauthorblockA{\IEEEauthorrefmark{1}Old Dominion University, Norfolk, VA, USA\\
\{snaru002, jrafieia, mghasemi, takabi\}@odu.edu}
\IEEEauthorblockA{\IEEEauthorrefmark{2}University of Arizona, Tucson, AZ, USA\\
eduardoblanco@arizona.edu}
\thanks{1. Both authors have equal contribution to this research.}
}

\maketitle
\begin{abstract}
\textbf{Large Language Models (LLMs)} remain vulnerable to \textbf{multi-turn jailbreak attacks}. We introduce \textbf{HarmNet}, a \textbf{modular framework} comprising \textbf{ThoughtNet}, a hierarchical semantic network; a \textbf{feedback-driven Simulator} for iterative query refinement; and a \textbf{Network Traverser} for real-time adaptive attack execution. HarmNet systematically explores and refines the adversarial space to uncover \textbf{stealthy, high-success attack paths}. Experiments across closed-source and open-source LLMs demonstrate that HarmNet outperforms \textbf{state-of-the-art} methods, achieving significantly \textbf{higher attack success rates}. For example, on Mistral-7B, HarmNet achieves a 99.4\% \textbf{attack success rate}—13.9\% higher than the best baseline.
\end{abstract}


\section{Introduction}
LLMs have demonstrated state-of-the-art performance across diverse applications such as education, healthcare, legal reasoning, and customer support \cite{zhao2023survey,hagos2024recent}. Despite extensive work on alignment—ranging from taxonomy of model risks \cite{weidinger2022taxonomy} and surveys of evaluation frameworks \cite{chang2024survey} to Reinforcement Learning from Human Feedback \cite{ouyang2022} and Constitutional AI \cite{bai2022}—LLMs remain susceptible to adversarial manipulation known as jailbreak attacks \cite{yi2024jailbreak}. These attacks craft malicious prompts that subvert safety filters, coaxing models into producing harmful or illicit content \cite{zou2023universal,wei2024}.

Early efforts focused on \emph{single-turn} jailbreaks, employing prompt-based adversarial tactics 
\cite{PromptAttack2023, PAIR2023, DrAttack2024}

More recently, \emph{multi-turn} strategies have leveraged conversational context to distribute malicious intent across seemingly benign exchanges. Crescendo iteratively escalates benign queries to harmful ones \cite{russinovich2024}, Chain of Attack (CoA) guides attacks via semantic-driven chains \cite{yang2024chain}, Derail Yourself uncovers self-discovered clues for multi-turn exploits \cite{ren2024derail}, JSP splits harmful questions into puzzle-like segments \cite{JSP2024}, and MRJ-Agent uses reinforcement learning for adaptive dialogue attacks \cite{MRJAgent2024}. Despite their advances, these methods either explore limited subspaces of the adversarial landscape or depend heavily on hand-crafted heuristics.

To address these limitations, we introduce \textbf{HarmNet}, a modular, LLM-agnostic framework that systematically constructs, refines, and executes multi-turn jailbreak queries. HarmNet comprises (1) \emph{ThoughtNet}, a hierarchical semantic network for comprehensive adversarial space exploration; (2) a \emph{feedback-driven Simulator} for iterative query refinement based on harmfulness and semantic alignment; and (3) a \emph{Network Traverser} for real-time adaptive attack execution. In our experiments on the HarmBench benchmark \cite{mazeika2024harmbench}, HarmNet achieves a 94.8\% attack success rate on GPT-4o—10.3\% higher than the best prior method—and 91.5\% on GPT-3.5 Turbo. It also obtains 98.4\% on LLaMA-3-8B \cite{dubey2407llama} and 99.4\% on Mistral-7B \cite{jiang2023mistral7b}, demonstrating its robustness across both closed-source and open-source LLMs.

\begin{table*}[htbp]
\caption{Attack Success Rate (ASR, \%) across Closed-Source and Open-Source LLMs}
\begin{center}
\begin{tabular}{|l|c|c|c|c|c|c|}
\hline
\textbf{Method} 
  & \multicolumn{3}{|c|}{\textbf{Closed-Source}} 
  & \multicolumn{3}{|c|}{\textbf{Open-Source}} \\
\cline{2-7}
  & GPT-3.5 & GPT-4o & Claude 3.5 
  & LLaMA 3-8B & Mistral 7B & Gemma 2-9B \\
\hline
GCG \cite{zou2023universal}            
  & 55.8 & 12.5 & 3.0  
  & 34.5 & 27.2 & 24.5 \\
\hline
PAIR \cite{PAIR2023}                   
  & 41.0 & 39.0 & 3.0  
  & 18.7 & 36.5 & 28.6 \\
\hline
CodeAttack \cite{jha2023codeattack}    
  & 67.0 & 70.5 & 39.5 
  & 46.0 & 66.0 & 54.8 \\
\hline
RACE \cite{ying2025reasoning}          
  & 80.0 & 82.8 & 58.0 
  & 75.5 & 78.0 & 74.5 \\
\hline
CoA \cite{yang2024chain}               
  & 16.8 & 17.5 & 3.4  
  & 25.5 & 18.8 & 19.2 \\
\hline
Crescendo \cite{russinovich2024}       
  & 48.0 & 46.0 & 50.0 
  & 60.0 & 62.0 & 12.0 \\
\hline
ActorAttack \cite{ren2024derail}       
  & 86.5 & 84.5 & 66.5 
  & 79.0 & 85.5 & 83.3 \\
\hline
\textbf{HarmNet (Ours)}                  
  & \textbf{91.5} & \textbf{94.8} & \textbf{68.6}
  & \textbf{98.4} & \textbf{99.4} & \textbf{99.6} \\
\hline
\end{tabular}
\label{tab:asr_full}
\end{center}
\end{table*}

\section{Our Approach}
In this section, we describe HarmNet through three components: (1) \emph{ThoughtNet Construction}, which builds a rich semantic network of candidate adversarial paths; (2) \emph{Feedback‐Driven Simulation}, which iteratively refines and prunes these paths using harmfulness and semantic benchmarks; and (3) \emph{Network Traverser}, which selects and executes the best multi‐turn chain in real time. We use simple notation and equations to make each step clear.

\subsection{ThoughtNet Construction}
We begin with a user’s harmful intent expressed as a prompt \(q\).  First, we extract its core goal \(g\) using a lightweight instruction‐tuned LLM.  Next, we generate a set of candidate topics
\[
\mathcal{Z} = \{z_1, z_2, \dots, z_N\}
\]
by prompting the LLM to propose semantically relevant concepts, subject to
\[
\cos\bigl(\mathbf{v}_{z_i}, \mathbf{v}_g\bigr) \;\ge\; \tau_z,
\]
where \(\mathbf{v}_{z_i}\) and \(\mathbf{v}_g\) are embedding vectors of the topic and goal, and \(\tau_z\) is a similarity threshold that ensures each topic meaningfully relates to \(g\).

For each topic \(z_i\), we generate a diverse set of contextual sentences
\[
\mathcal{S}_{z_i} = \{s_{i1}, s_{i2}, \dots, s_{iM}\}
\]
that satisfy both relevance and diversity constraints:
\[
\cos\bigl(\mathbf{v}_{s_{ij}}, \mathbf{v}_g\bigr) \;\ge\; \tau_s
\quad\text{and}\quad
\cos\bigl(\mathbf{v}_{s_{ij}}, \mathbf{v}_{s_{ik}}\bigr) \;<\; \tau_d
\quad (\forall\, j\neq k).
\]
Here, \(\tau_s\) enforces alignment with the goal and \(\tau_d\) prevents redundant samples.  We additionally link each sample \(s_{ij}\) to a small set of entities \(e_{ijk}\) drawn from predefined classes (e.g., Tools, Techniques, Regulations), grounding the network in actionable components.

Finally, for each triple \((z_i, s_{ij}, e_{ijk})\), we invoke a chain‐generation prompt to produce a short multi‐turn query chain
\[
\mathcal{C}_{ijk} = \{c_1, c_2, \dots, c_T\}.
\]
These chains incrementally steer a victim model from benign context toward the harmful intent \(g\).  Collectively, all generated chains form the initial \emph{ThoughtNet} search space:
\[
\mathcal{C} = \bigcup_{i,j,k} \mathcal{C}_{ijk}.
\]
This hierarchical, threshold‐driven construction ensures a semantically rich and dynamically expandable adversarial space without exhaustive enumeration.

\subsection{Feedback‐Driven Simulation}
To improve these candidate chains, we simulate multi‐turn interactions.  For each chain \(\mathcal{C}_{ijk}\) of length \(T\), let \(r_t\) be the victim model’s response to \(c_t\).  A judge model then assigns:
\[
H_t\in\{1,2,3,4,5\},\quad
S_t = \cos\bigl(\mathbf{v}_{r_t}, \mathbf{v}_g\bigr),
\]
where \(H_t\) measures harmfulness and \(S_t\) measures semantic alignment.  We compute marginal gains:
\[
\Delta H_t = H_t - H_{t-1}, 
\quad
\Delta S_t = S_t - S_{t-1}.
\]
If \(\Delta H_t < \mu\), the attacker LLM refines \(c_t\) to increase harmfulness; if \(\Delta S_t < \nu\), it refines \(c_t\) to improve relevance.  These refinements leverage both the sequence of prior responses and the judge’s explanatory feedback.

After each simulation pass, we prune any chain whose cumulative scores fall below minimum thresholds:
\[
\sum_{t=1}^{T} H_t < H_{\min}
\quad\text{or}\quad
\sum_{t=1}^{T} S_t < S_{\min}.
\]
This iterative refine‐and‐prune cycle converges on a compact set of high‐potential chains, focusing computational effort on the most promising adversarial paths.

\subsection{Network Traverser}
From the pruned set of chains, the Network Traverser selects the most effective sequence based on its simulated performance.  During real‐time execution, each query \(c_t\) is submitted to the victim model, and the judge immediately evaluates its response.  If the response achieves the maximum harmfulness score \(H_t = 5\), the attack is declared successful and traversal stops.  Otherwise, the attacker LLM is allowed a final, light refinement of \(c_t\) using the judge’s feedback before proceeding to \(c_{t+1}\).  This adaptive, turn‐by‐turn strategy ensures that HarmNet deploys the most effective multi‐turn jailbreak under real‐world latency and budget constraints.

\section{Experiments}

\subsection{Experimental Setup}
We evaluate HarmNet on both closed-source (GPT-3.5-Turbo, GPT-4o, Claude 3.5 Sonnet) and open-source (LLaMA-3-8B, Mistral-7B, Gemma-2-9B) LLMs using the HarmBench benchmark \cite{mazeika2024harmbench}. Baselines include single-turn methods (GCG \cite{zou2023universal}, PAIR \cite{PAIR2023}, CodeAttack \cite{jha2023codeattack}) and multi-turn methods (RACE \cite{ying2025reasoning}, CoA \cite{yang2024chain}, Crescendo \cite{russinovich2024}, ActorAttack \cite{ren2024derail}). GPT-4o serves as the attacker and, together with Flow-Judge, LLaMA-3-8B, and Mistral-7B, fulfills the judge role in simulation. Results are averaged over five independent runs to account for LLM stochasticity.

\subsection{Comparison with State-of-the-Art Attacks}
Table \ref{tab:asr_full} reports Attack Success Rates (ASR) across six target models. On closed-source models, HarmNet achieves 91.5\% on GPT-3.5-Turbo and 94.8\% on GPT-4o, surpassing ActorAttack by 5.0 and 10.3 percentage points respectively. On Claude 3.5 Sonnet, it attains 68.6\%, a 2.1-point gain over the next best. In the open-source setting, HarmNet markedly outperforms all baselines: 98.4\% on LLaMA-3-8B (vs.\ 79.0\% ActorAttack), 99.4\% on Mistral-7B (vs.\ 85.5\%), and 99.6\% on Gemma-2-9B (vs.\ 83.3\%). These improvements demonstrate that HarmNet’s structured exploration and feedback-driven refinement yield consistent, substantial gains over both heuristic and learning-based multi-turn attacks.

\subsection{Attack Diversity}
\begin{figure}[htbp]
  \centering
  \includegraphics[width=\columnwidth]{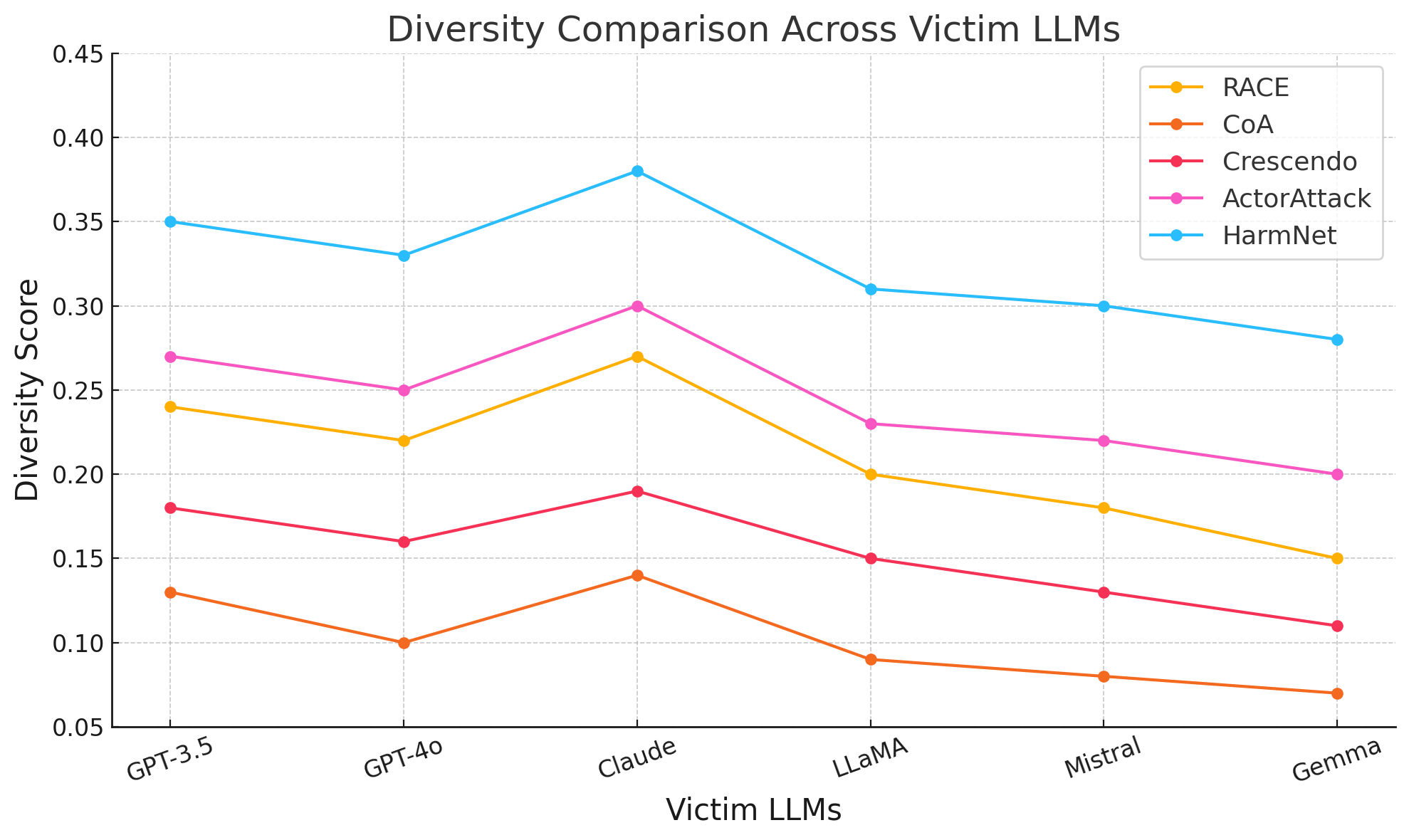}
  \caption{Attack Diversity Comparison}
  \label{fig:diversity_comparison}
\end{figure}

We also evaluate the \emph{semantic diversity} of full-dialogue prompts that achieve jailbreaks. Using MiniLMv2 embeddings \cite{wang2020minilmv2}, we compute pairwise cosine distances among concatenated multi-turn dialogues and define a diversity score:
\[
\mathrm{Diversity} = 1 \;-\; \frac{1}{|S|^2}\sum_{i>j}
\frac{\phi(x_i)\!\cdot\!\phi(x_j)}{\|\phi(x_i)\|\|\phi(x_j)\|},
\]
where \(S\) is the set of successful dialogues and \(\phi(\cdot)\) the embedding. As shown in Fig.~\ref{fig:diversity_comparison}, HarmNet consistently yields the highest scores (e.g., +15–25 points over ActorAttack), indicating it uncovers a broader range of adversarial trajectories. High diversity is critical for robust red-teaming, ensuring coverage of varied attack modes.

\end{document}